\documentclass{article}
\usepackage{times}

\usepackage{amsmath,amssymb}

\usepackage{url}

\begin{document}
%
\title{Operator Products in the SU($\infty$) Principal Chiral Model\thanks{Contribution to the proceedings volume for {\em Gravity, Strings and Fields: A conference in honor of Gordon Semenoff.}}}
%
%
\author{Peter Orland,\\ Baruch College, CUNY, 17 Lexington Av, 10010 USA and The Graduate School \\and University Center, CUNY, 365 Fifth Avenue,
New York, NY 10016, USA\\
orland@nbi.dk, peter.orland@baruch.cuny.edu}
%
%
%


\maketitle              

\begin{abstract}
The SU($N$) principal chiral model is asymptotically free and integrable in $1+1$ dimensions. In the large-$N$ limit, there is no scattering, but  correlation 
functions are {\em not} those of a free field theory. We briefly review the derivation of 
form factors for local operators. Two-point functions for such operators are known 
exactly. The two-point function of scaling-field operators has the short-distance behavior expected from the renormalization group.
We briefly discuss non-vacuum operator products. The ultimate goal is to derive the Lagrangian field theory from this axiomatic quantum-field-theory formalism.
\end{abstract}
\section{The Principal Chiral Model}

Although it is a pleasure to present this work to celebrate Gordon Semenoff's seventieth birthday, I have an ulterior motive. Problems remain
in the subject, calling out for more expertise and physical intuition: Gordon, help!

The SU($N$) principal chiral model (PCM) \cite{gursey} in 1+1 dimensions is considered a signpost for quantum chromodynamics. 
The PCM is an asymptotically-free matrix quantum field theory. It is known to have a mass gap, but this is not analytically proved. Integrability yields the exact S matrix for all $N$ \cite{exactSmatrix}. 



One method to study integrable quantum models is the form factor program, dating from the 1970's \cite{early}, formalized 
by Smirnov \cite{smirnov} (see Reference \cite{Bab} for a clear introduction). Unfortunately, finding many 
form factors of the PCM is hard, due to bound-state poles. The large-$N$ method is also limited. At large-$N$, the O($N$) and 
${\mathbb C}{\mathbb P}(N\!\!-\!\!1)$ models
have bubble diagrams, with a linear structure; but Feynman diagrams of the large-$N$ PCM are 
planar. {\em Combining} the large-$N$ and form-factor approaches, however, yields 
results \cite{summing1}, \cite{summing2}, \cite{ACC}.  


Form factors of local operators and exact expressions for correlation functions have been found. The 
short-distance two-point function of the scaling field  \cite{AsFr}, \cite{Katzav}, matches the renormalization group result \cite{RossiVicari}. The short-distance forms of other quantities are known, although the derivations are not yet published. 

These methods lie in the realm of axiomatic field theory. We hope to use the operator-product expansion to address the most pregnant question raised by this work: whether there is a proof of equivalence with Lagrangian field theory. If so, the model would be truly solved at large-$N$, firmly establishing the existence of the mass gap. This is tough, Gordon, and I need your guidance.

\subsection{The Basic Features of the PCM}

The Lagrangian of the SU($N$) PCM is, with ultraviolet cut-off $\Lambda$ implicit,
\begin{eqnarray}{\mathcal L}=\frac{N}{2g_{0}^{2}}{\rm Tr} \,\partial_{\mu}U^{\dagger}\partial^{\mu}U. \label{Lagrangian} 
\end{eqnarray} 
Here $g_{0}$ is the bare coupling in the 't$\,$Hooft limit $N\rightarrow \infty$. The global symmetry is $U(x)\rightarrow V_{L}U(x)V_{R}$, where
$V_{L}, V_{R}\in {\rm SU}(N)$ are the left- and right-transformations, respectively. Local operators include
the scaling field, $\Phi(x)$, its Hermitian conjugate $\Phi(x)^{\dagger}$, the left and right Noether currents $j^{\rm L}_{\mu}(x)$, $j^{\rm R}_{\mu}(x)$, respectively, and the stress-energy-momentum tensor (or stress tensor) $T_{\mu\nu}(x)$.

The scaling field $\Phi(x)$ is a complex $N\times N$ matrix field 
related to the bare field by
\begin{eqnarray}
\langle ...\Phi(x)....\rangle=Z^{-1/2}\langle ...U(x)....\rangle,     \label{scaling}
\end{eqnarray}
where $Z$ is a $\Lambda$ -dependent real renormalization factor. The field $\Phi(x)^{\dagger}$ is
defined similarly. The scaling field is neither unitary nor of unit determinant, in general, but (\ref{scaling}) is well-defined as $\Lambda\rightarrow\infty$.

The currents are
\begin{eqnarray}
j^{\rm L}_{\mu}(x)_{b}={\rm i}{\rm Tr}\,\,t_{b} \, \partial_{\mu}U(x)\, U(x)^{\dagger}\;,\;\;
j^{\rm R}_{\mu}(x)_{b}={\rm i}{\rm Tr}\,\,t_{b} \, U(x)^{\dagger}\partial_{\mu}U(x) \;, \label{currents}
\end{eqnarray}
where the generators $t_{a},\;a=1,\dots,N^{2}-1$, of SU($N$) satisfy ${\rm Tr}t_{a}t_{b}=\delta_{ab}.$

The stress tensor $T_{\mu\nu}(x)$ is defined as symmetric $T_{\nu\mu}(x)=T_{\mu\nu}(x)$, conserved $\partial^{\mu}T_{\mu\nu}(x)=0$, and has vanishing vacuum expectation value.

The beta function and the anomalous dimension of the scaling field are, respectively,
\begin{eqnarray} 
\beta(g_{0})=\frac{\partial g_{0}^2}{\partial \ln \Lambda}=-\beta_{1}g_{0}^{4}+\cdots,\; \;\beta_{1}=\frac{1}{8\pi}, \nonumber 
\end{eqnarray}
\begin{eqnarray}
\gamma(g_{0})=\frac{\partial}{\partial \ln \Lambda}\ln \frac{1}{N}\langle {\rm Tr}\Phi(x) \Phi(0)^{\dagger}\rangle   =\gamma_{1}g_{0}^{2}+
\cdots,\;\; \gamma_{1}=\frac{N^2 -1}{4\pi N^2}. \nonumber  
\end{eqnarray}
The asymptotically-free short-distance behavior of the scaling field is
\begin{eqnarray}
\frac{1}{N}\langle {\rm Tr}\,\Phi(x) \Phi(0)^{\dagger}\rangle \simeq C \ln^{\gamma_{1}/\beta_{1} }(\vert mx\vert^{-1})
\longrightarrow C \ln^{2 }(m\vert x\vert). 
\label{short-distance}
\end{eqnarray}
where $m$ is the mass gap and the limit is taken as $N\rightarrow \infty$ \cite{RossiVicari}. We recovered this result from entirely nonperturbative methods \cite{AsFr}. With E. Katzav \cite{Katzav}, we found the universal constant $C$ in
(\ref{short-distance}),
which is impossible to calculate with perturbation theory. 

Form factors of currents and the stress tensor have also been found (including some form factors for currents for finite $N$) \cite{ACC}, \cite{ACCPO}, \cite{ACCThesis}. There were errors in Reference \cite{ACCPO}, some of which were corrected in \cite{ACCThesis}.

\subsection{The S Matrix and the 't$\,$Hooft limit}

We first discuss the S matrix of the chiral Gross-Neveu model with symmetry
SU($N$). The rapidity $\theta$ of a mass-$m$ excitation parametrizes energy and momentum
by, respectively, $m\cosh\theta_{j}=E_{j},\;m\sinh\theta_{j}=p_{j}$. Rapidity differences are denoted by $\theta_{jk}=\theta_{j}-\theta_{k}$. 

An S matrix with SU(N) invariance has asymptotic particle states, labeled by $r=1,\dots, N-1$ with corresponding masses obeying the 
sine law (this is a general kinematic 
feature of
integrable quantum field theories \cite{OldSineGordon})
$
\mu_{r}=\mu_{1} \sin\frac{\pi r}{N}/\sin\frac{\pi}{N}.$
These are bound states of $r$ elementary particles (whose label is $r=1$). The $r=N-1$ case is the antiparticle of the $r=1$ particle. The 
chiral-Gross-Neveu S-matrix element of two such elementary particles \cite{BKKW}, each with a single color quantum number $a,c$ is
\begin{eqnarray}
_{\rm out}\langle p^{\prime}_{1},a^{\prime};p^{\prime}_{2}c^{\prime}\vert p_{1},a;p_{2},c\rangle_{\rm in}
= \langle a^{\prime},c^{\prime}\vert S_{\rm CGN}(\theta)\vert a,c\rangle  \delta(p^{\prime}_{1}-p_{1}) \delta(p^{\prime}_{2}-p_{2})    , \label{forward}
\end{eqnarray}
where
\begin{eqnarray}
S_{\rm CGN}(\theta)\!\!=\!\!\frac{\Gamma({\rm i}\theta/2\pi+1)
\Gamma(-{\rm i}\theta/2\pi-1/N) }{\Gamma({\rm i}\theta/2\pi+1-1/N) \Gamma(-{\rm i}\theta/2\pi)}
\left({  1\!{\rm l} 
}-\frac{2\pi{\rm i}}{N\theta}P\right), \label{CGNSmatrix}
\end{eqnarray}
where $1\!{\rm l}$ keeps colors the same in the initial and final states, $P$ exchanges the two initial colors in the final state, and where $\theta=\theta_{12}$. There is only forward scattering in (\ref{forward}). The bound-state poles of this S matrix have residues of order $1/N$ \cite{KS}. 

The mass spectrum of the PCM also obeys a sine law
$
m_{r}=m_{1} \sin\frac{\pi r}{N}/\sin\frac{\pi}{N}.
$
The rules for the spectrum are similar to the chiral-Gross-Neveu case. The main difference is that the excitations are color dipoles. Again, there 
is only forward scattering. The S matrix of 
two elementary ($r=1$)
particles contains two factors of (\ref{CGNSmatrix}) \cite{exactSmatrix}:
\begin{eqnarray}
S_{\rm PP}(\theta) =\frac{\sin (\theta/2-\pi{\rm i}/N)}{\sin(\theta/2+\pi{\rm i}/N)}\;S_{\rm CGN}(\theta)\otimes 
S_{\rm CGN}(\theta)
\label{PCMSmatrix}
\end{eqnarray}

Note that the binding energy of bound states with $r=2,\dots, N-2$ vanish in the 't Hooft limit. As is the case for the chiral Gross-Neveu model, the residues of
the bound-state poles vanish. Thus only the elementary particle ($r=1$) and antiparticle ($r=N-1$) survive as $N\rightarrow \infty$. Consequently, for the rest of
our analysis, we need only (\ref{PCMSmatrix}) and its crossed version $S_{PA}(\theta)$, representing the scattering of an elementary antiparticle 
particle with an elementary particle. Crossing is implemented by exchanging the s- and t-channels, and 
replacing $\theta$ by ${\hat \theta}=\pi {\rm i}-\theta$. The $1/N$-expansion of the particle-particle S-matrix element (\ref{PCMSmatrix}) 
is 
\begin{eqnarray}
S_{\rm PP}(\theta) \!\!= \!\!\left[1+O\left(\frac{1}{N^{2}}\right)\right] \!\!\! \left[ 1\!{\rm l} \!\otimes \!1\!{\rm l}-\frac{2\pi {\rm i}}{N\theta}\ P \!\otimes \!1\!{\rm l}
-\frac{2\pi {\rm i}}{N\theta} 1\!{\rm l}\!\otimes \!P -\frac{4\pi^{2}}{N^{2}\theta^{2}} P\!\otimes \!P \right]. \label{1/N-exp}
\end{eqnarray}

The scattering matrix of one particle and one antiparticle $S_{PA}(\theta)$ is obtained by crossing
(\ref{1/N-exp}) from the $s$-channel to the $t$-channel:
\begin{eqnarray}
&\!\!S_{PA}\!\!&(\theta)^{d_{2}c_{2};c_{1}d_{1}}_{a_{1}b_{1};b_{2}a_{2}}=\left[ 1+O(1/N^{2})\right] \left[\delta^{d_{2}}_{b_{2}}\delta^{c_{2}}_{a_{2}}\delta^{c_{1}}_{a_{1}}\delta^{d_{1}}_{b_{1}}\right.
\nonumber \\
&\!\!\!-\!\!\!&\!\!\!\left. \frac{2\pi{\rm i}}{N{\hat \theta}}
\!\left(\! \delta_{a_{1}a_{2}}\delta^{c_{1}c_{2}}\delta^{d_{2}}_{b_{2}}\delta^{d_{1}}_{b_{1}}
+\delta^{c_{2}}_{a_{2}}\delta^{c_{1}}_{a_{1}}\delta_{b_{1}b_{2}}\delta^{d_{1}d_{2}}
\! \right)\! -\!\!\frac{4\pi^{2}}{N^{2}{\hat \theta}^{2}}
\delta_{a_{1}a_{2}}\delta^{c_{1}c_{2}}\delta_{b_{1}b_{2}}\delta^{d_{1}d_{2}} 
\right]\!\!. \label{crossed-S}
\end{eqnarray}


\section{Large-$N$ Form Factors of the PCM}

\subsection{Rules for Exact Form Factors}

The form factor of a local operator ${\mathfrak B}(x)$ is an expression of the form
given by 
\begin{eqnarray}
f^{\mathfrak B}(\theta_{1},q_{1}; \dots; \theta_{n},q_{n})=\langle 0\vert {\mathfrak B}(0) \vert \theta_{1},q_{1};\dots;\theta_{n},q_{n}\rangle_{{\rm in}},
\label{matrix-element}
\end{eqnarray}
where $\theta_{1},\dots,\theta_{n}$ are the external rapidities of excitations and $q_{1},\dots q_{n}$ are the corresponding quantum numbers. Other 
matrix elements, besides (\ref{matrix-element}), may be found by crossing particle states from the ket to the bra.

To determine form factors, it is convenient to use Smirnov's axioms \cite{smirnov}, \cite{Bab}:

%
%

\noindent
{\em 1. Scattering Axiom}
\begin{eqnarray}
&\!\!f\!\!&\!\!^{\mathfrak B}(\theta_{1},q_{1};\dots ;\theta_{i-1},q_{i-1};\theta_{i+1},q_{i+1};\theta_{i},q_{i};
\theta_{i+2},q_{i+2};\dots; \theta_{n},q_{n})   \nonumber \\
&\!\!=\!\!& 
S^{k_{i} k_{i+1}}_{q_{i} q_{i+1}}(\theta_{i}-\theta_{i+1})
f^{\mathfrak B}(\theta_{1},q_{1} ;\dots;\theta_{i-1},q_{i-1};\theta_{i},q_{i};
\theta_{i+1},q_{i+1};
\theta_{i+2},q_{i+2}\dots;\theta_{n},q_{n}) .   
\;\;\;\;\;\;\;\;\;\;\;
\nonumber 
\end{eqnarray}
This axiom is a generalization of Watson's theorem.

%
%
\noindent
{\em 2. Periodicity Axiom}
\begin{eqnarray}
f^{\mathfrak B}(\theta_{1},q_{1};\dots;\theta_{n},q_{n})
=f^{\mathfrak B}(\theta_{n}-2\pi {\rm i},q_{n};\theta_{1},q_{1};\dots;\theta_{n-1},q_{n-1})\;.
\nonumber 
\end{eqnarray}

%
%

\noindent
{\em 3. Annihilation-Pole Axiom}
\begin{eqnarray}
&\!\!{\rm i}\!\!&\!\!\left. {\rm Res} f^{\mathfrak B}(\theta_{1},q_{1};\dots ;\theta_{n},q_{n})\right\vert_{\theta_{n}=\theta_{n-1}+\pi{\rm i}}
=f^{\mathfrak B}(\theta_{1},q_{1};\dots,\theta_{n-2};q_{n-2}) \,C_{q_{n-1} q_{n}} \nonumber \\
&-&  
S^{k_{n-1} k_{1}}_{t_{1}q_{1}}(\theta_{1}-\theta_{n-1})
S^{t_{1} k_{2}}_{t_{2}q_{2}}(\theta_{2}-\theta_{n-1})\cdots
S^{t_{n-4} k_{n-3}}_{t_{n-3}q_{n-3}}(\theta_{n-3}-\theta_{n-1})   \nonumber
\\
&\times & S^{k_{n-3} k_{n-2}}_{q_{n-1}q_{n-2}}(\theta_{n-2}-\theta_{n-1})   
f^{\mathfrak B}(\theta_{1},k_{1};\dots ;\theta_{n-2},k_{n-2})\,C_{k_{n-1} q_{n}} \;, 
\nonumber 
\end{eqnarray}
where $C$ is the charge-conjugation matrix. The normalization of the left-hand side
follows from the standard state normalization, {\em e.g.} $<\!\!\theta\vert\theta^{\prime}\!\!>
=4\pi\delta(\theta-\theta^{\prime})$. This axiom and the previous one are special cases of ``generalized crossing", discussed in detail
by Babujian et.al. \cite{Bab}. See also Appendix B in Reference \cite{smirnov}.


\noindent
{\em 4. Lorentz-Invariance Axiom}
If an operator ${\mathfrak B}(x)$ carries Lorentz spin s (not the center-of-mass energy
squared $s$), the form
factors must
transform under a boost $\theta_{j}\rightarrow \theta_{j}+\alpha$ for all $j=1,\dots, n$ as
\begin{eqnarray}
f^{\mathfrak B}(\theta_{1}+\alpha, q_{1};\dots; \theta_{n}+\alpha,q_{n})=e^{{\rm s}\alpha}
f^{\mathfrak B}(\theta_{1},q_{1}; \dots; \theta_{n},q_{n}) \;, \nonumber 
\end{eqnarray}


\noindent
{\em 5. Bound-State Pole Axiom} 
The name says it all. Poles appear in the form factors, at differences of rapidities corresponding to the formation of bound states.


\noindent
{\em 6. Minimality Axiom} 
In the spirit of the bootstrap, we assume form factors have as much analyticity as possible. It can sometimes be checked
that this is consistent with some perturbative method, which means either standard covariant
perturbation theory or
the $1/N$-expansion. In cases where we can find form factors, minimality stands up very well. If
a first guess for the form factor $f^{\mathfrak B}(\theta_{1}, \dots, \theta_{n})$ satisfies
the first four axioms, then so does
\begin{eqnarray}
f_{\rm minimal}^{\mathfrak B}(\theta_{1}, \dots, \theta_{n})=
f^{\mathfrak B}(\theta_{1}, \dots, \theta_{n})
\frac{P_{n}(\{ \cosh(\theta_{j}-\theta_{k}) \})}{Q_{n}( \{  \cosh(\theta_{j}-\theta_{k}) \} )}\;,
\nonumber 
\end{eqnarray}
where $P_{n}$ and $Q_{n}$ are symmetric polynomials. In this way, we can eliminate all the
poles in form factors, except those required by Axiom 3. or
corresponding to bound states. For Axiom 3
to be satisfied:
\begin{eqnarray}
\left.P_{n} \right\vert_{\theta_{n}=\theta_{n-1}+\pi{\rm i}}=P_{n-2}\;,\;\;
\left.Q_{n} \right\vert_{\theta_{n}=\theta_{n-1}+\pi{\rm i}}=Q_{n-2}\; . \nonumber
\end{eqnarray}

The overall normalization of form factors can often be found through physical or mathematical considerations.

Although the S matrix becomes unity as $N\rightarrow\infty$, the correlation functions are not those of a free field, as is clear from (\ref{short-distance}). Technically, the nontriviality of our form factors is due to the fact that factors of $1/N$ in the particle-antiparticle S-matrix elements (\ref{crossed-S}) are cancelled by
traces appearing in the application of Axiom 2. above.

\subsection{The Scaling Field of the SU($\infty$) PCM}


We use the shorthand 
$P_{j}=P,\theta_{j}, a_{j}, b_{j}$, for particle states and $A_{j}=A,\theta_{j},b_{j},a_{j}$ for antiparticle states (note that we have reversed the order of the left color and the right color in this last expression). The form factors were found by applying the scattering axiom to the first external excitation (with rapidity $\theta_{1}$) repeatedly, pushing this excitation past
all the others, finally comparing with the periodicity axiom \cite{summing1}, \cite{summing2}. 
The 
form factor of the scaling field, for an $M-1$-antiparticle and $M$-particle in-state is \cite{summing1}, \cite{summing2}
\begin{eqnarray}
\langle\!\!\!\!\!&\!\!\!\!\!\!0\!\!\!\!\! \!\!&\!\!\!\!\!\vert \Phi(0)_{b_{0}a_{0}}\;\vert A_{1};\dots;A_{M-1}; P_{M};\dots ;P_{2M-1}\rangle 
\nonumber \\
&\!\!\!=\!\!\!&\!\!\!
N^{-M+1/2}\sum_{\sigma,\tau\in S_{M}}
F_{\sigma \tau}(\theta_{1},\theta_{2},\dots,\theta_{2M-1})
\prod_{j=0}^{M-1}\delta_{a_{j}\;a_{\sigma(j)+M}} 
\delta_{b_{j}\;b_{\tau(j)+M}}\; ,
\label{special-FF}
\end{eqnarray}
where $S_{M}$ is the permutation group of $M$ objects,  and $F_{\sigma\tau}=F^{0}_{\sigma\tau}+O\left(\frac{1}{N}\right)$, with
\begin{eqnarray}
F^{0}_{\sigma \tau}(\theta_{1},\theta_{2},\dots,\theta_{2M-1})
=
\frac{ (-4\pi)^{M-1}   \prod_{l=1}^{M} \left[ 1-\delta_{\sigma(l)\tau(l)}    \right] 
}
{\prod_{j=1}^{M-1} 
[\theta_{j}-\theta_{\sigma(j)+M}+\pi{\rm i}][\theta_{j}-\theta_{\tau(j)+M}+
\pi{\rm i}]}. \label{MFF}
\end{eqnarray}
The expression (\ref{MFF}) has no double poles. The simple poles
occur where differences of rapidities are equal to $\pm \pi {\rm i}$. The annihilation-pole axiom was used to find the overall normalization. 

Substituting the $M=2$ form factor into the LSZ reduction formula yields the $1/N$ expansion of the 
two-particle S matrix \cite{summing1}. 


\subsection{Current Operators}

Dropping the superscript $L$ from the left current (\ref{currents}), its form factor for one particle and one antiparticle \cite{ACC} is 
\begin{eqnarray}
\langle 0\vert j_{\mu}(0)_{a_{3}a_{0}}\vert P_{1},A_{2} \rangle_{\rm in} 
&=&\frac{2\pi {\rm i}(p_{1}-p_{2})_{\mu}}{\theta_{12}+\pi {\rm i}}
\delta_{a_{3}a_{1}} \delta_{a_{0},a_{2}}\delta_{b_{1}b_{2}} \nonumber \\
&=&\frac{2\pi {\rm i} \epsilon_{\mu\nu}\eta^{\nu\lambda} (p_{1}+p_{2})_{\lambda}}{\theta_{12}+\pi {\rm i}}\tanh\frac{\theta_{12}}{2}
\delta_{a_{3}a_{1}} \delta_{a_{0},a_{2}}\delta_{b_{1}b_{2}}\,.  \label{lowestCFF}
\end{eqnarray}
The first expression on the right-hand side clearly satisfies the Lorentz-invariance axiom. The latter form is more convenient for
checking the annihilation-pole axiom of form factors with more excitations. The normalization is fixed by requiring the charge operator
$Q_{0}=\int j_{0} dx$ to satisfy the SU($N$) commutation relations \cite{ACC}.

The generalization to the form factor with $M$ antiparticles and $M$ particles, $M=1,2\dots$, in the in-state was found by A. Cort{\'e}s Cubero
\cite{ACCThesis}:
\begin{eqnarray}
\langle \!\!\!\!&\!\!\!\!0\!\!\!\!&\!\!\!\!\vert j_{\mu}(0)_{a_{2M+1}a_{o}}\vert P_{1};\dots;P_{M};A_{M+1};\dots ;A_{2M} \rangle_{\rm in}
=\frac{(4\pi)^{M}{\rm i}{\tilde P}_{\mu}}{2N^{M-1}} \nonumber \\
&\times\!\!\!&\!\!\!\sum_{\stackrel{\sigma\in S_{M+1}}{ \tau\in S_{M}}} \left[ \prod_{l=0}^{M} \delta_{a_{l}\;a_{M+\sigma(l)}+1}\right]
\left[ \prod_{l=0}^{M} \delta_{b_{l}\;b_{M+\tau(l)}}\right]J_{\sigma\tau}(\theta_{1},\dots,\theta_{2M}), \label{CFF}
\end{eqnarray}
where ${\tilde P}_{\mu}=\epsilon_{\mu\nu}\eta^{\nu\alpha}(p_{1}+\cdots+p_{2M})$ and $J_{\sigma\tau}=J^{0}_{\sigma\tau}+O\left(\frac{1}{N}\right)$, with
\begin{eqnarray}
\!\!\!\!\!&\!\!J^{0}\!\!&\!\!_{\sigma\tau}(\theta_{1},\dots,\theta_{2M})=
\tanh \frac{\theta_{M+1+\sigma(0)} -\theta_{\sigma^{-1}(M)} }{2} \nonumber \\
\!\!\!&\!\!\!\times\!\!\!&\!\!\frac{   \prod_{l=1}^{M} \left[ 1-\delta_{1+\sigma(l)\;\tau(l)}    \right] 
}
{ \left\{ \prod_{j=1}^{M}\left[ 1-\delta_{\sigma(j)\;M} \right]
[\theta_{j}-\theta_{\sigma(j)+M+1}+\pi{\rm i}] \right\}\prod_{j=1}^{M}[\theta_{j}-\theta_{\tau(j)+M}+
\pi{\rm i}]}.\label{Jfactor}
\end{eqnarray}

\subsection{The stress tensor}

The large-$N$ stress tensor form factors can be found using the same methods. The result is
\begin{eqnarray}
\langle \!\!\!\!&\!\!\!\!0\!\!\!\!&\!\!\!\! \vert T_{\mu\nu}(0)\vert P_{1};\dots;P_{M};A_{M+1};\dots ;A_{2M} \rangle_{\rm in}
=-\frac{2^{M-1}\pi^{M+1}{\tilde P}_{\mu}  {\tilde P}_{\nu}   }{N^{M-1}} \nonumber \\
&\times\!\!\!&\!\!\!\sum_{\sigma, \tau\in S_{M}} \left[ \prod_{l=0}^{M} \delta_{a_{l}\;a_{M+\sigma(l)}}\right]
\left[ \prod_{l=0}^{M} \delta_{b_{l}\;b_{M+\sigma(l)}}\right]T_{\sigma\tau}(\theta_{1},\dots,\theta_{2M}), \label{EMTFF}
\end{eqnarray}
where $T_{\sigma\tau}=T^{0}_{\sigma\tau}+O\left(\frac{1}{N}\right)$, with
\begin{eqnarray}
T^{0}_{\sigma\tau}
\!\!=\!\!
\tanh^{2} \frac{\theta_{2M}-\theta_{1} }{2} 
 \prod_{l=1}^{M}\frac{   1-\delta_{\sigma(l)\;\tau(l)}     
}
{  
[\theta_{j}-\theta_{\sigma(j)\!+\!M}\!+\!\pi{\rm i}] [\theta_{j}\!-\!\theta_{\tau(j)\!+\!M}+
\pi{\rm i}]}.\label{Tfactor}
\end{eqnarray}

\section{Correlation Functions}

\subsection{Closed-Form Expressions}

The Wightman function of the scaling field can be written as a series by expanding in terms of a complete set of states 
\begin{eqnarray}
{\mathcal W}_{N}(x)
=\frac{1}{N}\langle 0 \vert {\rm Tr}\Phi(0) \Phi(x)^{\dagger}\vert 0\rangle
=\frac{1}{N}\sum_{a_{0},b_{0}}\sum_{X}\,
\left\vert \langle 0 \vert \Phi(0)_{b_{0}a_{0}} \vert X\rangle_{\rm in}\right\vert^{2}\,
e^{{\rm i}p_{X}\cdot x} , \nonumber 
\end{eqnarray}
where $X$ denotes an arbitrary choice of particles, momenta and colors and where $p_{X}$ is the 
momentum eigenvalue of the state $\vert X\rangle$. This series is, for $N\rightarrow\infty$, explicitly (here we have reordered the numbering of the rapidities)
\begin{eqnarray}
\!\!\!\!\!\!\!{\mathcal W}(x)\!\!=\!\!\!\int \frac{d\theta}{4\pi} e^{{\rm i}x\cdot p} \!\!+\!\!\frac{1}{4\pi}\sum_{l=1}^{\infty} \!\!\int d\theta_{1}\cdots d\theta_{2l+1}
e^{{\rm i}x\cdot (p_{1}+\cdots p_{2l+1} \!)}\!\!
\prod_{j=1}^{2l}\frac{1}{(\theta_{j}-\theta_{j+1})^{2}+\pi^{2}},
\label{series}
\end{eqnarray}
where $l=M-1$. For further details on how (\ref{series}) was obtained, see Reference \cite{summing2}.

We can find a similar result for the current-current correlation function \cite{ACCThesis},
\begin{eqnarray}
\lim_{N\rightarrow\infty} \frac{1}{N}\langle \!\!\!\!&\!\!\!\!\!\!0\!\!\!\!\!\!&\!\!\!\!\!\!\vert j^{a}_{\mu}(0)j^{b}_{\nu}(x)\vert 0\rangle
=-\frac{\delta^{ab}\eta_{\mu\nu}}{4}
\sum_{M=1}^{\infty}\int d\theta_{1}\cdots d\theta_{2M} e^{{\rm i}x\cdot (p_{1}+\cdots p_{2\!M})} \nonumber \\
&\times\!\!&\!\!(p_{1}+\cdots+p_{2M})^{2} \tanh^{2}\frac{\theta_{2M}-\theta_{1}}{2}
\prod_{l=1}^{2M-1} \frac{1}{(\theta_{l}-\theta_{l+1})^{2}+\pi^{2}}.
\end{eqnarray}

\subsection{Short-Distance Asymptotics}

The older short-distance method exploits a Wick rotation to Euclidean space \cite{cardyetc.}. Here is a 
Minkowski-space version. Defining $\zeta=\tanh^{-1}x^{1}/x^{0}$, we have 
\begin{eqnarray}
x\cdot p_{j}=R \cosh(\theta_{j}-\zeta),\;\;R=\sqrt{x\cdot x}. \nonumber
\end{eqnarray}
Lorentz boosting by $\zeta$, changes $\theta_{j}\rightarrow \theta_{j}+\zeta$, $x^{0}\rightarrow R$, $x^{1}\rightarrow 0$, and
\begin{eqnarray}
e^{{\rm i}x\cdot (p_{1}+\cdots p_{2l+1} \!)} \rightarrow \exp \left[ {\rm i} mR(\cosh \theta_{1}+\cdots +\cosh \theta_{2l+1}) \right]. \nonumber
\end{eqnarray}
For small $R$, the {\em distribution} $\exp mR\cosh\theta$ is asymptotically $1$ for $\vert \theta \vert <L\equiv \vert \ln(mR)\vert$, and $0$ for
$\vert \theta \vert >L$ (where
this function oscillates rapidly). Writing $u_{j}=\theta_{j}/L$,
\begin{eqnarray}
W(x)
=\frac{L}{2\pi}\!+\!\! \frac{L}{4\pi}\sum_{l=1}^{\infty} \int_{-1}^{1} \!\!du_{1}\cdots \int_{-1}^{1} \!\!du_{2l+1}\;
\prod_{j=1}^{2l}\frac{1}{L[(u_{j}-u_{j+1})^{2}+(\pi/L)^{2}]}
.\label{Gseries1}
\end{eqnarray}

Recently we learned that
Kac and Pollard \cite{KP} found the same results for expressions like (\ref{Gseries1}) as we had \cite{AsFr}, \cite{Katzav}. Like us, they summed 
 (\ref{Gseries1}) using properties of the
square root of the Laplacian on functions $f(u)$, with $f(\pm1)=0$ \cite{MRiesz}:
\begin{eqnarray}
{\Delta}^{1/2}f(u) =\sqrt{-\frac{d^{2}}{du^{2}}}=\frac{1}{\pi}\int_{-1}^{1}du^{\prime}\;{\rm PV}\;\frac{f(u^{\prime})-f(u)}{(u^{\prime}-u)^{2}}\;, \label{FL}
\end{eqnarray}
where PV denotes the principal value. The Green's function is \cite{MRiesz}, \cite{KP}, \cite{BGR}
\begin{eqnarray}
K(u,v)=\Delta^{-1/2}\delta(u-v)=\frac{1}{2\pi}\ln\frac{ 1-uv+(1-u^{2})^{1/2}(1-v^{2})^{1/2} }{  1-uv-(1-u^{2})^{1/2}(1-v^{2})^{1/2}}. \label{Green}
\end{eqnarray}
Now \cite{AsFr}, \cite{Katzav}, \cite{KP}
\begin{eqnarray}
\frac{1}{L[(u-u^{\prime})^{2}+(\pi/L)^{2}]}
=
\langle u^{\prime} \vert e^{-\frac{\pi}{L}   {\Delta}^{1/2}+O(1/L)  }
\vert u \rangle. \label{TM}
\end{eqnarray}
Defining $\Delta^{1/2}\phi_{n}(u)=\lambda_{n}\phi_{n}(u)$, $u\in(-1,1)$, we can sum the series (\ref{Gseries1}):
\begin{eqnarray}
W(x)
=\frac{L}{4\pi}\sum_{n} \left\vert  \int_{-1}^{1}du\; \varphi_{n}(u,L)      \right\vert^{2} \frac{1}{1-e^{-2\pi\lambda_{n}/L+O(1/L^{2})}}
\simeq CL^{2}, 
\label{resum} 
\end{eqnarray}
for small $L$, where 
\begin{eqnarray}
C=\frac{1}{8\pi^{2}}\sum_{n}\left\vert  \int_{-1}^{1}du\; \varphi_{n}(u)      \right\vert^{2} \lambda_{n}^{-1} =\frac{1}{16\pi}. \label{coeff}
\end{eqnarray}
agreeing with (\ref{short-distance}). Our nonperturbative calculation has revealed the asymptotic freedom of the PCM. This is the only method yielding the coefficient
$C$ \cite{Katzav}. 

In work to appear, we found the short-distance forms for correlation functions of currents and the stress tensor. For example, the former is
\begin{eqnarray}
\frac{1}{N}\langle 0 \vert j_{\mu}^{a}(0)j_{\nu}^{b}(x)\vert 0\rangle \simeq \frac{\eta_{\mu\nu}\delta^{ab}}{16x^{\mu}x_{\mu}}, \label{currentCF}
\end{eqnarray}
which is consistent with currents having no anomalous dimensions.

\section{General matrix elements of operator products}

The work in this section describes an open problem. We wish to make the connection between our results and Lagrangian field theory. For example, (\ref{CFF}),
(\ref{Jfactor}) imply that currents are conserved: $\partial^{\mu}j_{\mu}=0.$ The remaining classical equation of motion is $[\partial_{0}+j_{0},\partial_{1}+j_{1}]=0$. We would expect to recover a corrected version of this equation of motion. Another problem is 
to find the effective coupling $g(R)$, where again $R=\sqrt{x\cdot x}$, we would need to check each side of
\begin{eqnarray}
T_{\mu\nu}(0)=\frac{g(R)^{2}}{N} \left[j_{\mu}\!\!\left(-\frac{x}{2}\right)j_{\nu}\!\!\left(\frac{x}{2}\right)-\frac{1}{2}\eta_{\mu\nu} j^{\alpha}\!\!\left(-\frac{x}{2}\right)j_{\alpha}\!\!\left(\frac{x}{2}\right)\right]
\end{eqnarray}
The left-hand side has known matrix elements, e.g., 
\begin{eqnarray}\langle P\vert T_{\mu\nu}(0) \vert P^{\prime}\rangle=-\frac{2\pi^{2}}{(2\pi{\rm i}+\theta_{PP^{\prime}})^{2}}(p+p^{\prime})_{\mu}(p+p^{\prime})_{\nu}\delta_{aa^{\prime}}
\delta_{bb^{\prime}}.
\end{eqnarray}
To solve either of these problems requires finding non-vacuum matrix elements of bilinears of current operators, that is
\begin{eqnarray}
_{\rm in}\langle E_{1},\dots,E_{m}\vert j_{\mu}(0)j_{\nu}(x)\vert E^{\prime}_{1},\dots,E^{\prime}_{n}\rangle_{\rm in}. \label{bilinear}
\end{eqnarray}
These can be written as integrals, just as the vacuum expectation values can. The integrals are harder to evaluate, however. For large $L$, with
$\tanh\zeta=x^{1}/x^{0}$, as before,
\begin{eqnarray}
\!\!\!\!\!\!\!\!\!\!\!\!\!\!\!\!\!\!\!&\langle \!\!\!&\!\!\!P\vert[ j_{\mu}(x)j_{\nu}(y)]_{a_{1}c_{1}} \vert P^{\prime}\rangle=\delta_{a_{1}a}\delta_{c_{1}a^{\prime} }\delta_{bb^{\prime}}
\frac{\partial^{2}}{\partial x^{\mu}\partial y^{\nu}}\frac{1}{2}\int_{-1}^{1}du\int_{-1}^{1}dv   K(u,v)\nonumber \\
\!\!\!&\!\!\!\times\!\!\!&\!\!\!
\coth \!\frac{Lu\!-\!\theta\!-\!\zeta}{2}\coth \!\frac{Lu\!-\!\theta^{\prime}\!-\!\zeta}{2}
\frac{1}{v-(\theta+\zeta-2\pi{\rm i})\!/\!L} \frac{1}{v-(\theta^{\prime}+\zeta+2\pi{\rm i})\!/\!L},
\end{eqnarray}
where $K(u,v)$ is the Green's function (\ref{Green}). We have similar results for multi-excitation states (\ref{bilinear}), but have not yet been able to simplify such expressions.

%

\paragraph{Acknowledgement:}
This work was supported in part by a grant from the PSC-CUNY.
%
%

\end{document}